# Saharan mineral dust outbreaks observed over the North Atlantic island of La Palma in summertime between 1984 and 2012

Benjamin A. Laken[1, 2], Hannu Parviainen[1, 2], Enric Pallé[1, 2] and Tariq Shahbaz[1, 2]

[1] Instituto de Astrofísica de Canarias, Via Lactea s/n, E-38205, La Laguna, Tenerife, Spain

[2] Department of Astrophysics, Faculty of Physics, Universidad de La Laguna, Tenerife, Spain

* Corresponding author: Benjamin A. Laken, blaken@iac.es

We estimate the frequency of Saharan mineral dust outbreak events observed over the North Atlantic island of La Palma based on *in situ* nightly atmospheric extinction measurements recorded almost continuously since 1984 by the Carlsberg Meridian Telescope at the Roque de los Muchachos observatory. The outbreak frequency shows a well-defined seasonal peak in the months of July to September, during which time the occurrence of Saharan dust events (SDEs) is approximately 28±6%. We find considerable year-to-year variability in the summertime SDEs frequency, observing a steady reduction between 1984 and 1997, followed by a period of relative mean stability from 1999 to 2012. We investigated changes in the atmospheric extinction of the SDEs as an indicator of strength of the episodes and found that this parameter approximately follows the SDE frequency, however, instrumental limitations prevented us from deriving precise conclusions regarding their long-term changes. A lagged correlation analysis between SDE properties and the El Niño Southern Oscillation (ENSO), North Atlantic Oscillation (NAO), and Sahel rainfall index (SRI) was performed. We found that 55±4% of the year-to-year variations in July–September SDE frequency may be reproduced by a lagged relationship to the NAO conditions during the preceding October–December period, and 45±4% may be reproduced by a negative correlation to the SRI during the preceding February–April period. Based on these relationships it may be possible to obtain an approximate indication of the strength of the upcoming summertime dust season over the North Atlantic around half a year in advance.

## 1. Introduction

Mineral dust entrained and suspended in the deep Saharan planetary boundary layer (PBL) is periodically transported over great distances in discrete outbreak events that may last from hours to several days (Knippertz and Todd, 2012). Entrained dust typically posses diameters (*d*) ranging from 0.5 to 75μm (Maring *et al*., 2002) and have been observed suspended at altitudes ranging from 4–6km above mean sea level (AMSL) (Esselborn *et al*., 2009): coarse dust particles (*d* > 30μm) rapidly settle, remaining close to their source, while small particles (*d* < 10μm) remain suspend for long periods of time and, as a result, may be transported over globally significant distances (Shao, 2000; Cuesta *et al*., 2009; Creamean *et al*., 2013). Away from the source regions, a progressive depletion of small particles occurs by the processes of dry deposition and by scavenging and washout from precipitation (Shao, 2000).

Mineral dust produces a range of significant direct and indirect impacts, including effects on climate and radiative balance (Ångström, 1930; Sassen *et al*., 2003; Forster *et al*., 2007), cloud properties and precipitation (Criado and Dorta, 2003; Richarsdon *et al*., 2007; Ansmann *et al*., 2008; Seifert *et al*., 2010; Creamean *et al*., 2013), marine and terrestrial ecosystems (Jickells *et al*., 1998; Shinn *et al*., 2000; Kaufman *et al*., 2005), and soil development (Yaalon and Ganor, 1973; Vine, 1987; Muhs *et al*., 1990; Menédez *et al*., 2007). In particular, due to the climatological significance of aerosols and the high contribution of mineral dust to the global net aerosol burden, significant effort has been focused on increasing the understanding of Saharan dust sources and long-range transport with the aim of reducing the associated uncertainty (Knippertz and Todd, 2012).

The path of Saharan mineral dust outbreaks shifts with the seasonal movements of the Intertropical Convergence Zone (ITCZ) (Jankowiak and Tanré, 1992; Swap *et al*., 1996; Moulin *et al.,* 1997; Goudie and Middleton, 2001). Entrained dust is seasonally transported along three main pathways (Goudie and Middleton, 2001; Ben-Ami *et al*., 2012): westward over the North Atlantic Ocean towards the Americas (Carlson and Prospero, 1972; Moulin *et al*., 1997; Baars *et al*., 2011; Creamean *et al*., 2013); northwards across the Mediterranean to Southern Europe (Loÿe -Pilot *et al*., 1986); and eastwards across the Mediterranean towards the Middle East (Herut and Krom, 1996; Ganor *et al*., 1991). The westward flow over the North Atlantic Ocean is the largest by volume: it is estimated that each year 240±80 × $10^9$ kg of mineral dust is transported over the North Atlantic (Kaufman *et al*., 2005), accounting for 30–50% of total dust output from the Sahara (Schütz *et al*., 1981; Goudie and Middleton, 2001). Consequently, the Canary Island archipelago, located 28.1ºN 15.4ºW, approximately 100km from the Western Sahara, frequently experiences the effects of dust-laden Saharan winds.

The largest and most frequent outbreak events over the North Atlantic occur during the Northern Hemisphere summer season, associated with strong convective disturbances over West Africa at latitudes of 15ºN–20ºN that move westward in association with easterly waves emerging from the African coast at intervals of 3–4 days (Goudie and Middleton, 2001). In summer months, the development of the Bermuda-Azores high (a.k.a. the North Atlantic subtropical high) is also important in drawing dust-carrying wind from the tropical North Atlantic to the subtropical region (Jickells *et al*., 1998; Goudie and Middleton, 2001). During this period the outbreak events most frequently occur in the free-troposphere, as dust-laden air from the deep

Saharan PBL is advected above the relatively colder and denser North Atlantic marine-boundary layer (MBL) airmass, located between the surface and ~2km AMSL (Cuesta *et al*., 2009). Conversely, in winter months, outbreaks in the free-troposphere are rare. Instead, infrequent outbreaks events occur mainly within the MBL as a result of anticyclonic activity over North Africa (Viana *et al*., 2002). In Section 3.3, we shall give a quantitative description of seasonal outbreak frequencies from *in situ* observations.

Over recent decades, the frequency of Saharan dust events (SDEs) has shown significant fluctuations in response to climatic factors such as drought, and anthropogenic impacts on marginal desert regions (Goudie and Middleton, 2001). A recent study shows that proxy data suggests the occurrence of a doubling of desert dust output over the $20^{th}$ century, with model simulations suggesting that the increased output results from a combination of climate drying, carbon dioxide fertilization, and land use changes (Mahowald *et al*., 2010). Several studies suggest that year-to-year variability in Saharan dust transport may be connected to synoptic weather patterns indicated by climate indices, for example: desert dust over the North Atlantic and the Mediterranean has demonstrated correlations to the North Atlantic Oscillation (NAO) (Moulin *et al*., 1997); additionally, dust observed at Barbados transported across the North Atlantic shows anticorrelations to precipitation anomalies in the Sahel region (Prospero and Nees, 1986; Prospero and Lamb, 2003; Mahowald *et al*., 2009). Consequently, we shall investigate the relationship of these parameters (and El Niño) to observed outbreak events from long-term observations of SDEs.

The objective of this paper is to investigate long-term changes in the frequency and intensity of SDEs using an approximately 30-year record of *in situ* nightly atmospheric extinction measurements taken at the Carlsberg Meridian Telescope (CMT), 28.46ºN, 17.53ºW, 2.4km AMSL, located at Roque de los Muchachos observatory at La Palma (Canary archipelago). From this data, we examine the long-term properties of SDEs observed over the Canary Islands, and the associated variations in the North Atlantic Oscillation, El Niño, and rainfall anomalies over the Sahel. The location of the Canary Islands, immediately to the west of North Africa means that the archipelago intersects the important westward transport route for Saharan mineral dust, consequently, long-term observations from the Canary Islands present an excellent opportunity to observe changes in Saharan dust transport over the Atlantic region.

## 2. Data

Atmospheric extinction in optical wavelengths results from the absorption and scattering of incoming light as a result of water vapor, clouds, and aerosols (Varela *et al*., 2008). Nightly measurements of the atmospheric extinction coefficients have been recorded at the CMT almost continuously since 1984, the longest running dataset of its type. These measurements are centered on blue (551nm, Johnson’s V band, 13/05/1984–28/05/1998) and red (625nm, r’ band, 26/03/1999–present) wavelengths. The extinction coefficients are defined as the fractional depletion of radiance per unit path length. Specifically, the atmospheric extinction coefficient data measure the attenuation of photons (in magnitudes) as a function of wavelength, with distance defined in airmass units (mag. $airmass^{-1}$). As our data always concern values extrapolated to airmass 1, we shall refer to it only as the atmospheric extinction (K) with a unit of magnitude.

The atmospheric extinction measurements in the Johnson's V (551nm)-band at zenith (Johnson, 1963) were calculated from a nightly average of 56 photometric standard stars per night, as they crossed the north-south meridian (with a nightly standard deviation of ±38 stars over the 1984–2012 period) by a scanning slit micrometer; this method gave a single nightly atmospheric extinction value. From 26 March 1999 onwards, this instrument was replaced by a more sophisticated Charge Coupled Device (CCD) instrument, which measured atmospheric extinction in the Sloan Digital Sky Survey (SDSS) r' filter (625nm)-band, providing extinction data for each recorded frame. With this new instrument, nightly atmospheric extinctions are calculated from an average extinction value of all photometric images collected during a night. The instrument obtains extinction measurements by using an instrumental zero-point derived from the long-term performance of the CCD, with each image containing an average of 30–40 calibration stars. For the data to be judged photometric, it required a constant low degree of scattering (<0.15 magnitudes) between photometric standard stars (over the course of the observation night 1984–1998, and over each frame of data post-1999); where a minimum of ~10 standard stars are required to determine the scatter. During the post-1999 data collection period, photometric quality was also assessed based on the degree of scatter of the calculated extinctions between images over the course of the observation night; where the scatter was less than <0.06 magnitudes the night conditions and the data were classified as photometric. We note that the observations do not require completely clear sky conditions; if part of the sky is cloudy, observations may still be possible if enough standard stars can be observed. For further information regarding these atmospheric extinction measurement techniques see King (1985).

Previous studies have directly compared the atmospheric extinction data from the CMT to AERONET Sun photometer measurements of optical depth obtained from a station located approximately 200km to the east of the CMT, in Santa Cruz de Tenerife over a period of approximately 10 days (Bailey *et al*., 2008; Ulanowski *et al*., 2007). These studies found a good correspondence between the datasets during an outbreak of Saharan mineral dust over the Canary Islands, with a peak atmospheric extinction of ~0.30 observed by the CMT and ~0.26 by AERONET.

The sensitivity of atmospheric extinction measurements to the presence of mineral dust varies at different wavelengths. Consequently, we must apply a subtraction value to the V band data (0.0286) obtained from King (1985) to convert the atmospheric extinction data in the V band to the equivalent r' band values: these data are presented in Figure 1. Over the 1984–2012 period data coverage is 51%, with the majority of unobserved nights attributed to non-photometric (cloudy/windy/locally too humid) conditions. The number of photometric observation nights each year was comparable over the two data periods, with an average of 196±38 observations per year from 1984–1998, and 202±33 observations per year since 1999. The nightly atmospheric extinction data the mean and standard error uncertainties were 0.18±0.017 from 1984–1998 and 0.12±0.009 post-1999.

Several prominent gaps can be seen in the data, the most notable of these are a 301-day gap between 29 May 1998 and 26 March 1999, and a 145-day gap between 5

November 2007 and 29 of March 2008. The former resulted from the replacement of a scanning slit micrometer with a CCD camera (and the change from V band to r' band measurements), while the later resulted from a technical failure of the telescope.

We note that deriving the SDE properties from this data has limitations that will constrain the conclusions of our work, specifically these are: 1) The high-altitude location of the observatory (2.4km AMSL) means that we are limited to primarily observing SDEs occurring in the free-troposphere. During summertime, the Saharan planetary boundary layer adjacent to the Atlantic Ocean is usually exceptionally deep and developed, frequently achieving altitudes of up to 6km AMSL (Gamo, 1996; Esselborn *et al*., 2009). Consequently, during summertime outbreak events the mineral dust enriched air is normally transported in the free-troposphere above the relatively denser marine boundary layer (MBL) that extends to ~2km AMSL. The frequent advection of air from the Saharan PBL generates a layer referred to as the Saharan Air Layer (SAL) (Cuesta *et al*., 2009). However, between the months of January to March dust events occur in a very different manner: the outbreaks occur within the marine boundary layer (MBL) due to high-pressure systems affecting northern Africa (Alonso-Pérez *et al*., 2007). Since our observations are made at an altitude of 2.4km AMSL, we are certainly biased towards observing SDEs occurring in the free troposphere, and thus likely under-estimate the SDE occurrence in wintertime. 2) Our observations are restricted to photometric conditions only and to locally good observing conditions at the observatory site. As a result, if there is an association between SDE occurrence and cloud cover amount, then our analysis will under-estimate SDE occurrence. We note that this limitation is also true of aerosol data from AERONET and satellite-based observations. 3) A failure to note the presence of thin cirrus clouds may also result in an over-estimation of nightly atmospheric extinction values leading to an over-estimation of SDE occurrence. 4) Observations from the CMT are effectively a point-sample, restricted to the location above the observatory. Consequently, the data may fail to observe SDEs occurring in the region, which do not directly intersect the observatory: an example of such an event (27/02/2000) is noted by Varela *et al*. (2008).

## 3. Analysis

The following analysis shows how we developed the CMT atmospheric extinction data into a quantitative description of SDE frequency and strength (Sections 3.1 to 3.2); the seasonal and long-term properties of these parameters are investigated in Sections 3.3 to 3.4. An investigation of the correlation of these parameters against climate indices describing the North Atlantic Oscillation, the El Niño Southern Oscillation, and rainfall anomalies over the Sahel region is also presented in Section 3.5.

### *3.1. Atmospheric extinction 1984 – 2012*

Figure 1a shows both the CMT nightly atmospheric extinction values and, over plotted, a 100-day running median (red line). The majority of values occur at low atmospheric extinctions (<0.2 magnitudes) , in addition, groups of values with high atmospheric extinctions can be observed throughout the dataset, with the strength and number of events clearly showing year-to-year variability. We note two prominent features in Figure 1a: Firstly, the periodic occurrence of high atmospheric extinction values indicating a seasonal presence of Saharan dust outbreaks into the troposphere. Secondly, we note a period of several years following the eruption of Mount Pinatubo (June 1991) where an increase in the low-level atmospheric extinction values occurred as a result of volcanic aerosol injection into the stratosphere (Brock *et al.*, 1993), previously noted to affect atmospheric extinction from the CMT data (Sanroma *et al.*, 2010). In order to effectively isolate the effect of Saharan dust events (SDEs) on the atmospheric extinction values we have removed the 100-day running-median values from the nightly data. The median values act only as a low-pass filter, removing the effects of Pinatubo and other long-term variations from the data (where long-term refers to timescales significantly distinguishable from SDEs of >100 day). This has the effect of making the SDEs directly comparable from year-to-year: these results are presented in Figure 1b.

### *3.2 Period analysis*

To determine if any significant periodic variations occur in our data, we have applied the Lomb-Scargle (LS) period analysis method to the 5,352 unevenly sampled, clear-sky, nightly atmospheric extinction data of Figure 1b, and their standard errors (calculated from the scatter/square root of the number of images per night) (Scargle, 1982; Press *et al.*, 1992): the results are presented in Figure 2. The analyzed frequencies were constrained by the limits imposed by the Nyquist frequency and the duration of the dataset. The number of independent frequencies was determined using the method of Horne and Baliunas (1986). The standard procedure for detecting periodic features is based on estimating the noise spectrum and using this to define the point above which we are unlikely to observe a random fluctuation. The standard false alarm probability estimate from the LS algorithm gives the statistical significance of the highest peak in the power spectrum assuming that all data points are independent, however, in the presence of correlated data (i.e. red-noise), we will have to take a different approach in order to properly estimate the statistical significance of the peaks evidence in the periodogram. This was done numerically by means of Monte Carlo (MC) simulations. We generated data with exactly the same sampling as the real data with a modeled red-noise data generated using the method of Timmer and Knöig (1995) with a broken power law as determined from the periodogram of the observed data. We then added Gaussian noise using the uncertainties of the data (from the atmospheric extinction standard error). We then calculated the LS periodogram and recorded the position and frequency of the highest peak.

We computed 5,000 simulated datasets and calculated the 68$^{th}$ and 99.9$^{th}$ percentile confidence intervals at each frequency taking into account a realistic number of independent trials (*Vaughan*, 2005). From this method, we identified a statistically significant ($p < 0.001$) peak in atmospheric extinction at 366.01, 182.85 and 121.64 days, the first being an annual period while the later are aliasing frequencies. No other statistically significant periods were observed in the data, indicating that the remaining variability is not of a periodic nature.

### *3.3 Seasonal frequency and magnitude of Saharan Dust Events*

Presenting the atmospheric extinction data as a climatological average by calendar day clearly demonstrates seasonal characteristics, showing that the frequency and intensity of events greatly increases in boreal summer months (Figure 3). This is as expected, as the location of Saharan dust loading over the North Atlantic Ocean varies spatially with the seasonal movement of the Intertropical Convergence Zone (ICTZ) (Moulin *et al.*, 1997). However, we again note that our experimental sensitivity to dust outbreaks in boreal winter months is likely reduced due to the altitude at which our observations are made; this factor also likely contributes to the observed seasonal amplitude.

It is clear that the number of days with SDE conditions is far lower than with non-SDE conditions. In order to continue an analysis of long-term variations in the frequency and intensity of SDEs we must empirically distinguish these populations, defining a threshold value in atmospheric extinction above which we define the date as a SDE. We do this independently for the 1984–1998 V band values and post-1999 r' band values, as it is clear from an examination of Figure 1b that despite adjustment of the V band values, these differing measurement techniques still possess different sensitivities, with lower atmospheric extinction values consistently detected after 1999 in the r' band data.

Figure 4 presents two density plots of atmospheric extinction values following the removal of 100-day running median values to remove low-frequency variability. A normal-gamma composite distribution is used to model the population of the extinction values. The distribution is a linear combination of a normal distribution, modeling the extinction variability in non-dusty conditions. A gamma distribution is used to model the long, positive, tail caused by the SDE conditions. Namely, the composite distribution for atmospheric extinction (K) is $P(K; \mu, \sigma, o_g, a_g, b_g, c)$, where $\mu$ and $\sigma$ are the mean and standard deviation of the normal distribution, $o_g$, $a_g$, and $b_g$ are the origin, shape and width of the gamma distribution, and c is the mixing factor.

We fit the theoretical distribution to the observed atmospheric extinction values by using a Bayesian approach, and we compute the posterior probability distributions for the composite distribution parameters using Markov Chain Monte Carlo (MCMC). Uninformative constant priors are used on all the distribution parameters, and the likelihood is expressed directly as the product of the probabilities for obtaining each atmospheric extinction value from the composite distribution. Based on the mean and standard deviation posteriors of the normal-component of the composite distribution, we identify the 99.9$^{th}$ percentile values within the normal distribution as the point at which we consider all subsequent values to be outside the range of normal extinction variability: these atmospheric extinction values are 0.0522 and 0.0352 for the periods of 1984–1998 and 1998–2012 respectively.

We note that several earlier studies have similarly quantitatively distinguished non-dusty from dusty atmospheric extinctions over the Canary Island astronomical observation sites at both La Palma and Tenerife, identifying a range of values including: ≥0.153 (Guerrero *et al*., 1998); >0.20 (Siher *et al*., 2004); >0.075 (Jiménez *et al*., 1998); and >0.155 (García-Gil *et al*., 2010). However, the quantitative definitions of dusty conditions in these studies lacked a consideration of low-frequency variations in atmospheric extinction. Consequently, the ability of the various threshold values to correctly identify dusty conditions varies over time, which would make long-term analysis problematic. We further note that the thresholds defined are also sensitive to the wavelength with which atmospheric extinction is measured, which differed from study to study; this effect is evident in Figure 4, which shows differing SDE threshold values despite the application of a subtraction value to transform between the V and r' wavelengths.

We wish to examine two parameters relating to SDEs: firstly, the frequency with which SDEs occur over a specified period, defined here as the number of observed nights with an atmospheric extinction value above the SDE threshold values divided by the total number of observed nights (as a ratio of 1). Secondly, we investigate the intensity of SDEs, defined here by the mean atmospheric extinction value of observed SDE nights over a specified period (i.e. the average atmospheric extinction of nights with an extinction value of ≥ SED threshold values).

Regarding the error values presented in this work, unless otherwise stated all values indicate the ±1.96σ confidence level. Error values presented for the SDE frequency are based on Bayesian methods, whereby binomial distributions are calculated at each time step based on the total number of observed nights over a specified period and the number of nights with an atmospheric extinction of ≥ SED threshold values over the same period.

Plots of the monthly SDE frequency and atmospheric extinction climatologies are presented in Figure 5. A period of high SDE occurrence takes place during the months of July–September (hereafter also referred to as the high dust season). During this time SDE frequency is found to be 0.10±0.04 for all months excluding July–September, while, during the months of July–September the SDE frequency increases to 0.28±0.06 (Figure 5a). In Figure 5b we also present the month-to-month mean atmospheric extinction values during SDE conditions (a measure of SDE strength). We found the monthly variability to be relatively high, and despite the presence of a peak during the month of July, the values show a considerably weaker seasonality than the SDE frequency. A mean atmospheric extinction of 0.11±0.07 and 0.17±0.08 is observed in the months of October–June and July–September respectively (Figure 5b).

### *3.4. Changes in the frequency and strength of Saharan dust events during the high dust season from 1984 to 2012*

The number of clear-sky observations, SDE frequency, and the strength of SDEs during the high dust season between 1984 and 2012 are presented in Figure 6. Figure 6a displays the number of days with SDE conditions observed (dashed line) compared to the total number of observations (solid line) per-high dust season per-year. During 1998, no SDE conditions were observed, as the telescope was not operational during the change between measurement techniques.

To accurately gauge the uncertainty associated with the SDE frequency measurements, we have calculated a binomial probability density function for each annual dust season based on the number of SDEs and the number of total observations. The SDE frequency, along with ±1.96σ uncertainty values are presented in Figure 6b: we observe a general reduction in SDE frequency between 1984 to 1998 from 0.47±0.11 to 0.10±0.07, after which time values partially recovered. They have since remained in a state of relative long-term stability around a mean of 0.26±0.09 showing an average year-to-year σ of 0.11.

One of the largest variations in SDE frequency occurred in 1997, where, despite 60 nights of photometric observations, no SDEs were observed. This year is notable for the occurrence of a particularly strong El Niño event (Wolter and Timlin, 1998; McPhaden, 1999), suggesting that further investigation of the relationship between the El Niño Southern Oscillation (ENSO) and SDEs identified in this work are warranted: this will be addressed in the following section.

Figure 6c shows the atmospheric extinction (strength) of SDE events and $\pm 1.96\sigma$ error during the high dust season, and similarly identifies a long-term decline in SDE strength from 1984–1998, from $0.26 \pm 0.15$ to $0.09 \pm 0.04$. After 1999 SDE strength remained relatively stable (around $0.16 \pm 0.11$), with a mean year-to-year $\sigma$ of 0.04. We note that the higher atmospheric extinction values identified during the 1984–1998 period compared to the post-1999 period almost certainly result from the higher extinction values obtained by the V band period measurements previously noted. Consequently, we cannot reliably gauge the absolute atmospheric extinction changes over the entire data period, as there is no overlapping period within which we could calibrate the measurements. Despite this limitation, we may assume that the relative changes within the two distinct data collection periods (pre/post-1999) are reliable; i.e. a decrease in SDE strength occurred between 1984–1998, followed by a period of partial recovery and relative stability. We note that this limitation does not affect our ability to calculate SDE frequency over the entire data period, as this is a relative measurement, and therefore does not require consistent or calibrated measurement approaches over the complete duration of the observations.

### *3.5. Relationships to the NAO, MEI and SRI*

The correspondence between the SDE frequency minimum in 1997 and the large ENSO event of 1997–98 provides an indication that inter-annual variations in SDE frequency may be connected to large-scale climate oscillations. Indeed, correlations between inter-annual variations in dust transport and synoptic-scale climate indexes have been suggested by previous studies, e.g. year-to-year variations in Saharan dust export over the Atlantic and Mediterranean and the North Atlantic Oscillation (NAO) index (e.g. Moulin *et al*., 1997; Chiapelllo and Moulin, 2002; Dayan *et al*., 2008); correlations between winter-time dust export over the Canary Islands and the 1,000mb geopotential height anomalies between Tenerife and Madrid (Spain) (Alonso-Perez *et al*., 2011); and, Saharan dust export and drought severity over North Africa (Prospero and Nees, 1986; Prospero and Lamb 2003). Several studies have also identified relationships between specific synoptic meteorological conditions and dust emission from the Sahara (e.g. Klose *et al*., 2010; Knippertz and Todd, 2010; Alonso-Pérez *et al*., 2011).

From these studies, it seems likely that we may detect statistically significant relationships between regionally important climate parameters and SDE properties. Such relationships may provide further insights into the physical processes influencing SDE occurrence over the North Atlantic region. Consequently, we proceed by investigating the relationship between our observations and three climatological indexes suspected to be of significance: 1) the NAO index, defined as the difference in normalized sea level atmospheric pressures between the Azores high pressure region and the Icelandic low pressure region (Barnston and Livezey, 1987; Chen and van den Dool, 2003; van den Dool *et al*., 2000); 2) The Multivariate El Niño Southern Oscillation Index (MEI) (Wolter and Timlin, 1993; 1998), calculated from the first unrotated principle component of six combined variables observed over the tropical Pacific region (sea-level pressure, zonal and meridional surface winds, sea surface temperature, surface air temperature, and cloud fraction); and, 3) the Sahel Rainfall Index (SRI) (Janowiak, 1988), based on long-term precipitation measurements from stations within the National Center for Atmospheric Research (NCAR) World Monthly Surface Station Climatology (WMSSC) network within a region of 20ºN–8ºN, 20ºW–10ºE. The SRI data are presented in cm as an anomaly with respect to the period of 1950–1979.

A monthly-averaged time series of these three parameters from 1984–2012 are presented in Figure 7. Normally, the SRI shows a peak in precipitation during the wet season from the months of July–September, and low values during the dry season from the months of October–June. Since the values are presented as anomalies against the 1950–1979 mean, and overall precipitation has decreased since then, almost all the wet season peaks in our presented time interval occur as negative values. We note that the NAO index and SRI show a seasonality that we have removed prior to our correlation analysis; this was done by subtracting the monthly means of the 1984–2013 period from the individual monthly values.

To test for the presence of statistically significant relationships between these parameters and our SDE observations we have performed a lagged correlation analysis taking our observations of total counts (the total number of photometric nights), and SDE frequency as the dependent variables and the climate indexes averaged over 3-month periods as the independent variables. Using these data we performed a correlation analysis over a ±12 lag period, where each value represents a 3-month average beginning at a period denoted by the lag value: e.g. lag -1 are the months of June–August, lag 0 are the months of July–September, and lag +1 are the months of August–October. We note that all data from 1998 are removed from the dependent variables prior to the correlation analysis, as the minimum value of this year is an artifact. We reiterate that the dependent variables are only analyzed during the high dust season period (July–September): i.e. the analyzed time-series of the dependent variables is as presented in Figure 6a–b and does not change. Whereas, the independent datasets presented in Figure 7 are averaged (boxcar means) into three-month bins, and shift both forwards and backwards in time with respect to the dependent dataset. Thus, our analysis only concerns how the independent datasets are statistically related to the number of photometric observations at the CMT and the calculated SDE frequency during the high dust season.

The correlation coefficients (*r*-values) are obtained from linear regressions calculated using Markov Chain Monte Carlo (MCMC) techniques, which consider the error in both the independent and dependent variables where available. Thus, our presented correlations at each time point represent a distribution of ~40,000 probable *r*-values, for which we display the median (50$^{th}$ percentile) and ±1σ values. We evaluated the statistical significance of the MCMC-calculated correlations by means of a further series of MC simulations, wherein, the dependent variables were randomized 10,000 times and the correlations achieved were recorded for each independent/dependent variable pairing independently, from a distribution of the correlations we then extracted the 2.5$^{th}$/97.5$^{th}$ and 0.5$^{th}$/99.5$^{th}$ percentile correlations as the two-tailed 0.05 and 0.01 probability confidence intervals. This procedure was repeated at each time step, for each pair of variables (i.e. for each of the 10,000 randomly arranged dependent variables, the correlation at each lag time is calculated, with the resulting distribution of correlations determining the confidence intervals). The resulting confidence intervals were overplotted against the MCMC correlation values of each variable paring. We consider the resulting correlations statistically significant if the MCMC estimated median and ±1σ $r$-values simultaneously achieve $p$-values beyond the 0.05 two-tailed confidence interval. The results of this analysis are presented in Figure 8, and discussed in the following sections.

### *3.5.1. Correlations between the NAO and SDE properties*

The NAO bears a statistically significant positive correlation ($p$ <0.01) to the total number of observed nights at lag -1, with a median $r$-value of 0.65±0.17 (Figure 8a). The estimated lower 1σ $r$-value at lag -1 is statistically significant at $p$ <0.05, suggesting that the state of the NAO between the months of June–August directly influences the number of photometric nights (an indicator of weather conditions) between the months of July–September. This correlation returns approximately one year later, at lags of +11 to +12 months.

In relation to the NAO and SDE frequency we observe a highly statistically significant ($p$ < 0.01) positive correlation occurring at lags -10 to -8 (with a peak median $r$-value of 0.74±0.19): the significance of this case is so high that even the lower 1σ $r$-values are significant at the $p$ 0.01 level (Figure 8b). These correlations suggest a robust statistical link between the SDE frequency and the NAO during the preceding winter period (October–December period), implying that the NAO may account for 55±4% of the year-to-year variations in high dust season SDE frequency.

Our findings of significant correlations between the NAO and SDE properties are somewhat similar to observations of earlier studies such as Chiapello and Moulin (2002), who identified a correlation between optical thickness and the NAO during a twenty-year period over a region of the tropical North Atlantic, 15ºN–30ºN, 5ºW–30ºW (a region which includes the Canary Island archipelago). However, we note that the findings of Chiapello and Moulin (2002) pertain to winter dust transport and the winter NAO and so differ from our results. Similarly, Chiapello *et al.* (2005) also highlighted an influence of the NAO on the wintertime export of dust. These results also correspond to the findings of Siher *et al.* (2004), who found a positive correlation ($r$ 0.49) between the NAO and satellite-derived winter aerosol index data over a 10º × 10º area at zero-lag centered on La Palma between 1978–2002.

### *3.5.2. Correlations between MEI and SDE properties*

An analysis of the MEI and total observed nights show a positive correlation at the $p$ 0.01 confidence level at lag +4 (median $r$-value 0.47±0.19). The median significance of the MEI correlation slowly increases past the $p$ 0.05 level immediately following 0 lag, and remains between a $p$-value of 0.05 and 0.01 for 8 consecutive time-steps, although the lower 1σ level $r$-values remain at $p$ > 0.05 during the analysis period, indicating no statistically robust correlation is identified (Figure 8c). Similarly, the MEI also shows significant associations to SDE frequency, peaking at zero-lag (Figure 8d); however, the range of calculated $r$-values again suggests that this relationship is not robustly significant. The slow response of correlations to the MEI may result from a combination of: 1) the low month-to-month variability and consequent high degree of autocorrelation in the dataset (Figure 7b), and 2) an exaggeration of the autocorrelation effect by the boxcar mean approach of the lag periods. This result suggests that we do not identify a robust statistical link between ENSO and the local weather conditions over the observatory site or variations in the frequency of SDEs during the high dust season.

### *3.5.3. Correlations between the SRI and SDE properties*

The SRI shows strongly significant ($p < 0.01$) negative correlations with median *r*-values between lags -12 to -9 although the upper 1σ r-value only reaches statistical significance at lag -10 (Figure 8e); no other robustly significant relationships between the SRI and total observations are identified. Negative correlations between the SRI and SDE frequency are observed at several lags over the analysis period, however, the ±1σ *r*-value ranges for this case are not statistically significant at the *p* 0.05 level with the exception of lag -5, which possesses an *r*-value of -0.67±0.19 (Figure 8f). The confused nature of these and previously noted correlations is, at least in part, likely to be a consequence of the inter-relationships between these and further connected variables of the climate system.

We interpret these observations to suggest the following: Firstly, the negative relationship between the SRI and the total number of observations with a lag of -10 implies that large/small volumes of precipitation over the Sahel at the end of the wet season reduces/increases the number of photometric observation nights over La Palma. Secondly, the relationship between the SRI and the SDE frequency observed at lag -5 implies that increases/decreases in the volume of precipitation received during the dry season over the Sahel result in decreases/increases in the frequency of SDEs. In the following section we examine the correlations between the NAO/MEI and the SRI, as it is known that rainfall anomalies over the Sahel region are strongly related to sea surface temperature anomalies across the globe, e.g. Folland *et al.* (1986) and Hunt (2000).

### *3.6. Correlations between the SRI and NAO/MEI*

For almost all the NAO/MEI correlations to SRI the ±1σ *r*-value intervals are above the *p* 0.05 level, indicating the correlations are not statistically significant. There is one exception at lag +1 between the MEI and SRI, where the upper 1σ *r*-value is marginally below the *p* 0.05 significance level. This implies a statistically significant association between decreases/increases in the MEI and a subsequent increase/decrease in the SRI.

If ENSO were influencing the SDE frequency indirectly via a connection to the SRI, we should then expect to see a statistically significant positive correlation at lag -6 in the MEI–SDE frequency (Figure 8d). However, the only significant SRI–SDE frequency relationship detected in that case was a negative correlation at lag -5, suggesting that an indirect link between ESNO and SDE frequency via the SRI does not exist. Furthermore, both the correlations between the MEI–SRI and SRI–SDE frequency are negative: if the MEI–SDE relationship were being mediated by a mechanism related to precipitation over the Sahel, we would expect a positive correlation (at negative lags) between the MEI–SDE datasets, which is not observed.

### *3.7 Summary and discussion of the correlation analysis*

The correlation analysis has produced some ambiguous results, with numerous points of significance at both positive and negative lags. Consequently, we assessed the correlations in a conservative manner, only accepting results where the MCMC-calculated ±1σ distributions of *r*-values exceeded the two-tailed *p* 0.05 significance level. From this basis, we identified factors connected to variations in the high dust season SDE frequency: the NAO conditions from October–December, and the SRI conditions from February–April, respectively these correlations account for 55±4% and 45±4% of observed year-to-year variability. Although from a simple the summation of these $r^2$ values it would appear we may account for 100% of the variability in high dust season SDE frequency, this would only be true if the NAO and SRI indexes were completely independent, which of course is not the case.

Logically, we may consider that the observed SDE frequency may largely depend on: 1) The immediate weather of the North Atlantic/African region, which directly controls the entrainment and long-range transport of dust. 2) The dust entrainment and removal processes, which although are influenced by the immediate weather, may also be influenced by the amount of soil moisture available (to which the SRI is likely serving as a proxy). The precipitation and soil-moisture, in turn, depend on global-scale patterns of climate variability, and may involve appreciable lag-times and feedbacks to both weather and climate. Thus, a simple consideration of just these two factors, weather, and soil moisture, we have a considerable amount of inter-relationships and feedbacks that we cannot clearly resolve in a simple correlation analysis. Consequently, the results of our correlation-based examination is unavoidably limited by the complexity of the climate machine, and thus, likely to produce results which include ambiguous signals. As a result, while we are able to comment on statistical associations and their implications, climate model-based experiments are required to resolve this complexity, and to determine physical processes governing long-term SDE variations.

## 5. Conclusions

Utilizing approximately 30 years of nightly atmospheric extinction data recorded from the CMT located at Roques de los Muchachos of La Palma (Spain), we have examined year-to-year changes in the frequency and strength of Saharan mineral dust intrusions occurring between the months of peak dust activity (July–September). During this time, Saharan mineral dust is frequently transported in discrete outbreak events from the deep Saharan planetary boundary layer westwards, over the North Atlantic marine boundary layer in the free-troposphere. We observe a steady decline in SDE frequency from 1984–1997 from 0.47±0.11 to 0.10±0.07, after which time SDE frequency partially recovered and has fluctuated by approximately 0.11σ from a mean of 0.26±0.09. Although our conclusions regarding the long-term strength (atmospheric extinction) changes of the SDEs was limited by an inability to calibrate between periods of differing instrumentation, we note that it appears that the intensity of individual SDEs has approximately followed the variations observed in the SDE frequency.

A correlation analysis relating our observations to the North Atlantic Oscillation, the El Niño Southern Oscillation, and Saharan rainfall shows a variety of statistically significant relationships, although many of the correlations were ambiguous. Several correlations yield intriguing statistical associations, which imply mechanisms connected to the NAO and SRI influencing the year-to-year SDE frequency during the high dust season. Due to the time lags involved in these relationships, the NAO index values in December–April and SRI in February–April may be used as an indication of the strength of the next high dust season over the North Atlantic region.

## Acknowledgements

The authors would like to thank Daffyd Wyn Evans (University of Cambridge), Carsten Weidner (Instituto de Astrofísica de Canarias) and Beatriz Gonzalez-Merino (Instituto de Astrofísica de Canarias) for comments, and the reviews of Peter

Knippertz (University of Leeds) and a second anonymous referee. The MEI data were obtained from http://www.esrl.noaa.gov/psd/enso/mei, NAO index data were obtained from http://www.cpc.ncep.noaa.gov/data/teledoc/nao.shtml, and Saharan Rainfall Index data were obtained from http://jisao.washington.edu/data/sahel. The authors thank Daffyd Wyn Evans for providing the extinction data generated from observations carried out with the Carlsberg Meridian Telescope run by Real Instituto y Observatorio de la Armada en San Fernando, Copenhagen University Observatory and the Royal Greenwich Observatory. Hannu Parviainen received support from RoPACS during this research, a Marie Curie Initial Training Network funded by the European Commission's Seventh Framework program. Hannu Parviainen also received funding from the Väisälä Foundation through the Finnish Academy of Science and Letters during this research. Enric Pallé acknowledges support from the Spanish MICIIN, Grant No. #CGL2009-10641. Tariq Shahbaz acknowledges support from the Spanish Ministry of Economy and Competitiveness (MINECO) under the grant AYA2010-18080.

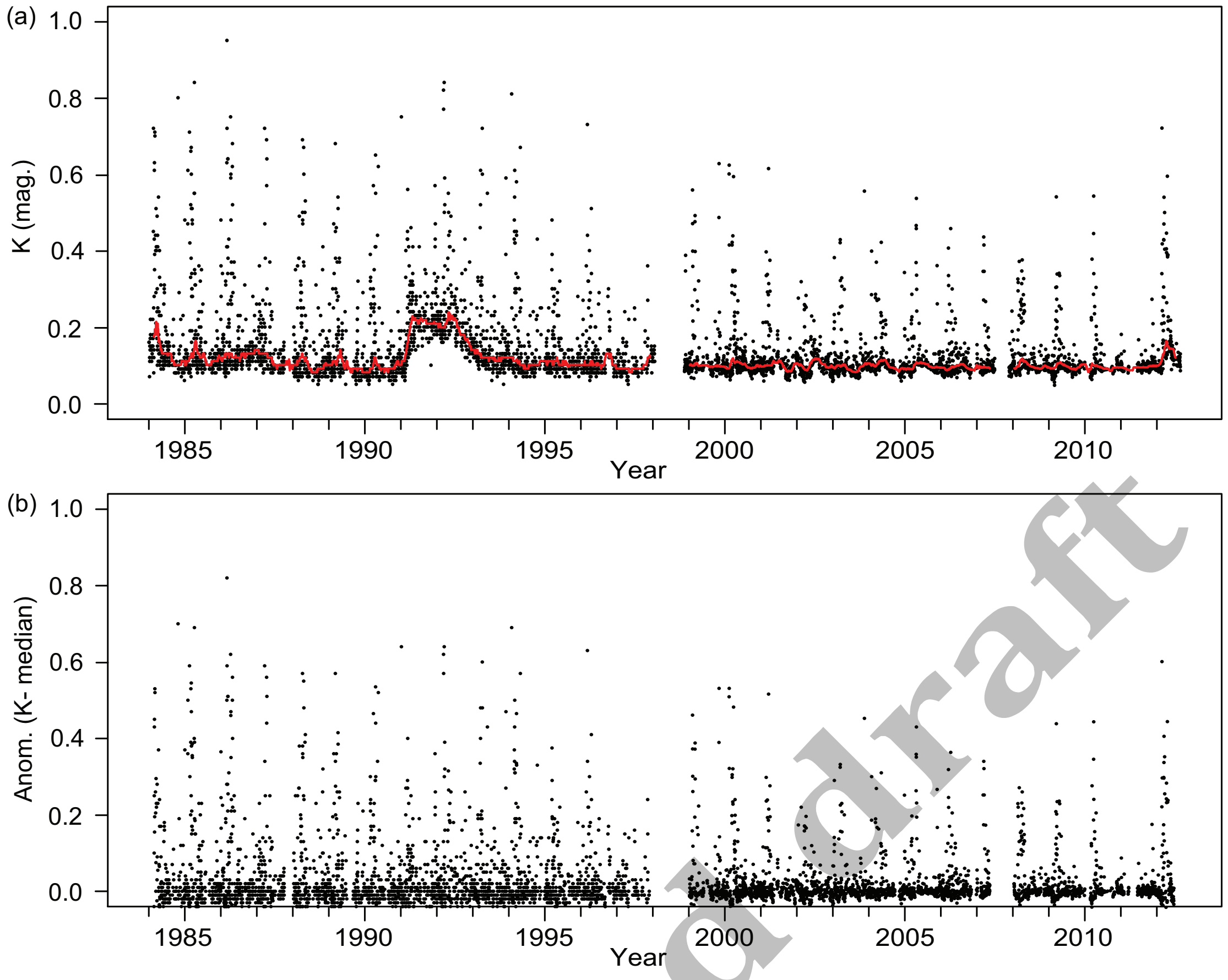

**Figure 1.** Time series of nightly atmospheric extinction (K, units in magnitudes) centered on the r' wavelength (625nm) from 13/05/1984–31/12/2012 for photometric nights, presented for: (a) raw data, with 100-day running median values (red line), and (b) the same data as an anomaly, after the median values are subtracted from each data point.

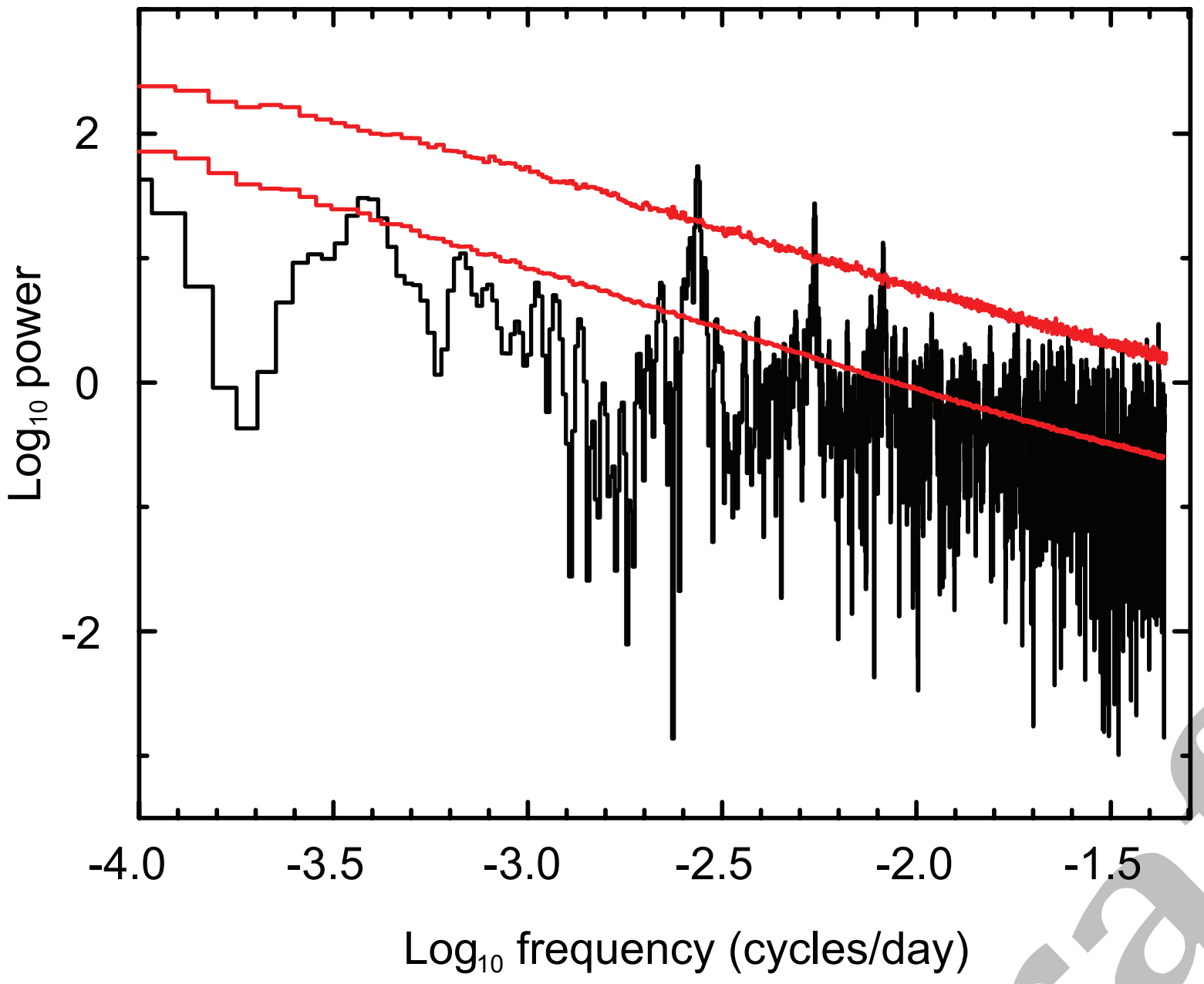

**Figure 2**. Lomb-Scargle period analysis of extinction data with 68th and 99.9th percentile two-tailed confidence levels, displayed on the red lines. Significant extinction peaks are observed at 121.64, 182.85, and 366.01 days; the later being an annual cycle, while the former are aliases of the annual cycle.

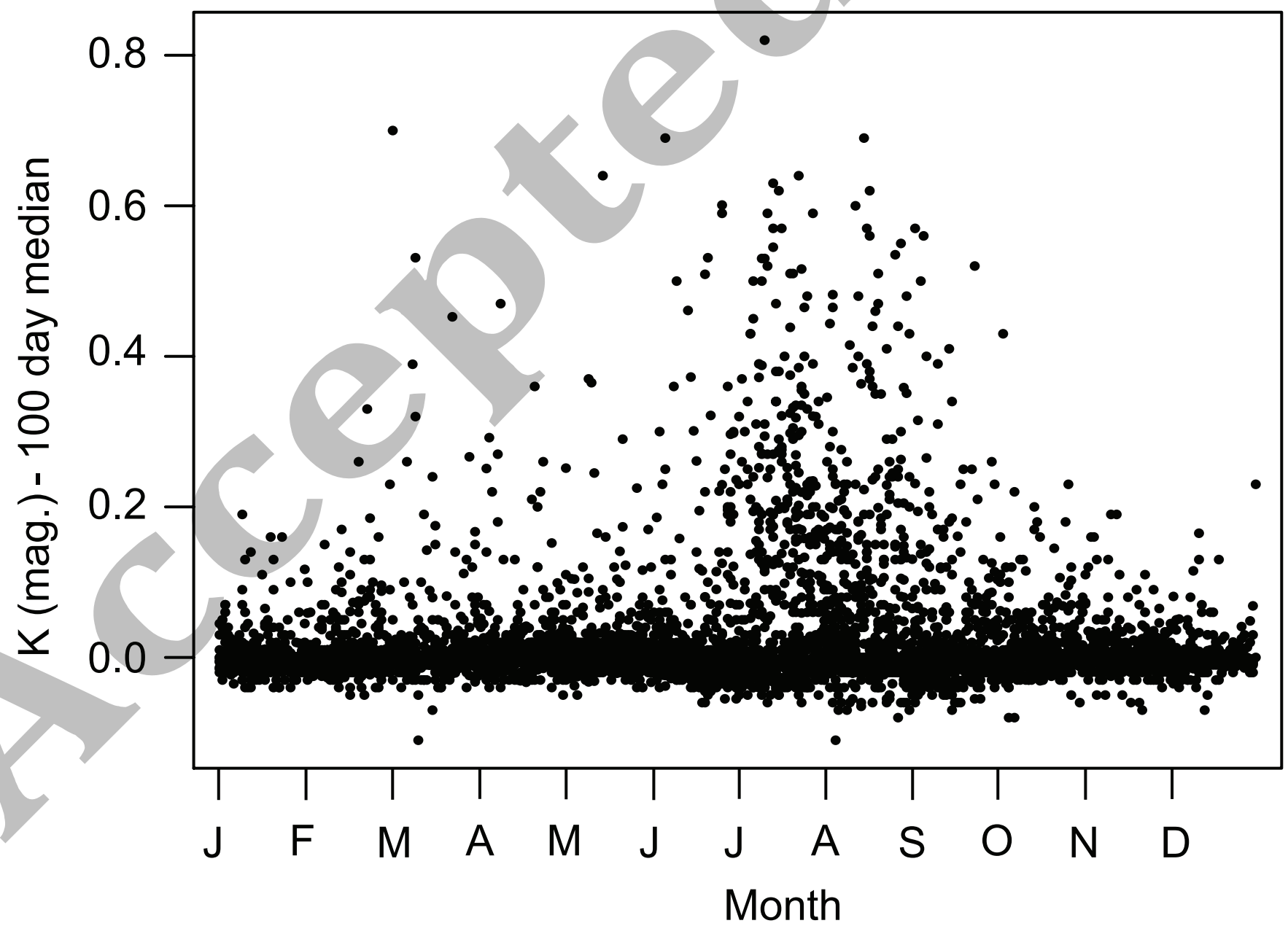

**Figure 3.** Scatter plot of clear-sky nightly atmospheric extinction (K) values observed since 1984 plotted seasonally (all values converted to equivalent r' band data by use of a subtraction value obtained from King *et al.*, (1985)), units in magnitudes.

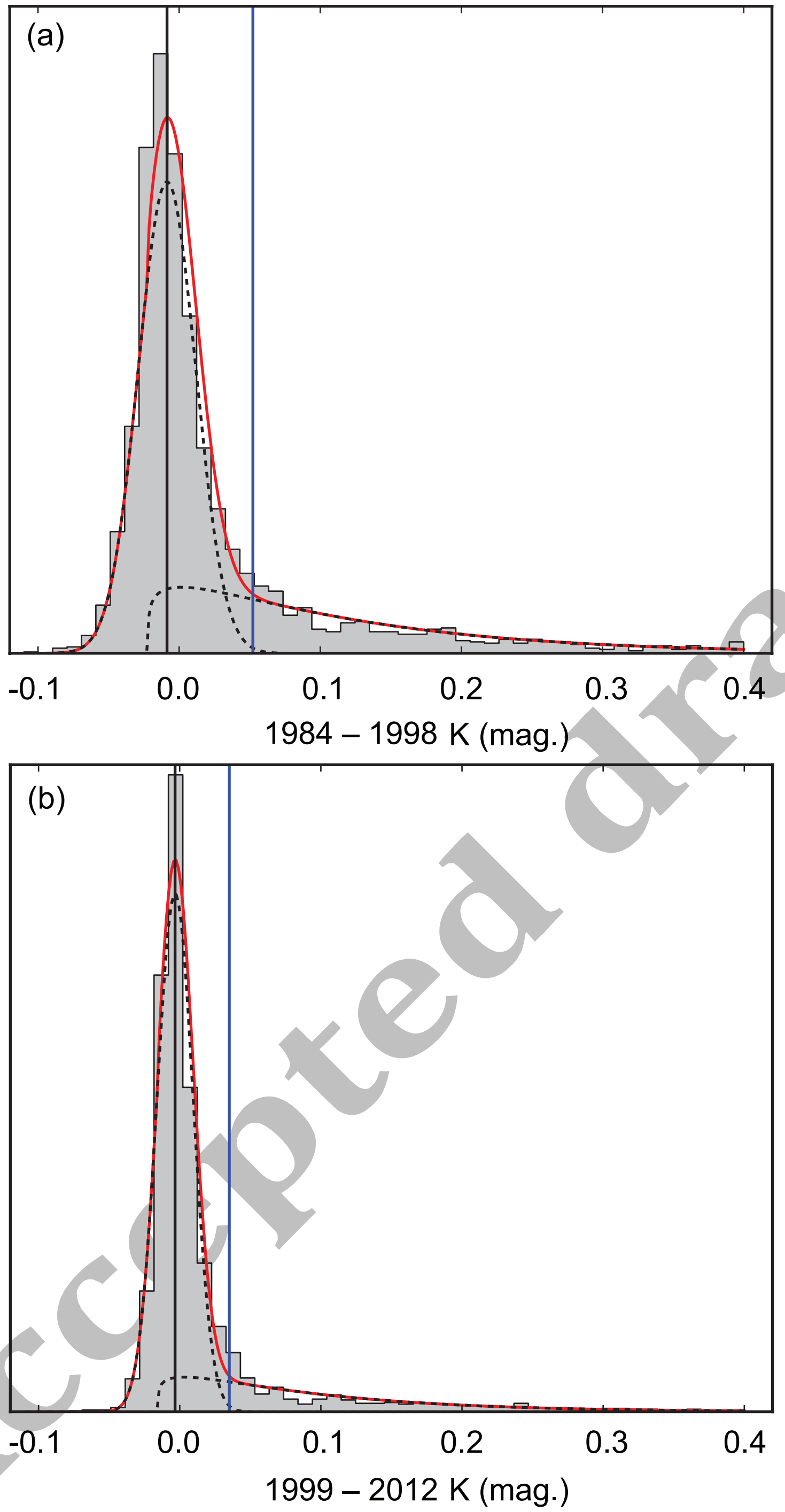

**Figure 4.** Density plot of atmospheric extinction (K, units in magnitudes) following removal of 100-day running medians. Markers indicate the mode (vertical black line) and Saharan dust-event (SDE) threshold values (blue line) found to be 0.0522 and 0.0352 respectively. The periods were treated independently due to their differing measurement sensitivities.

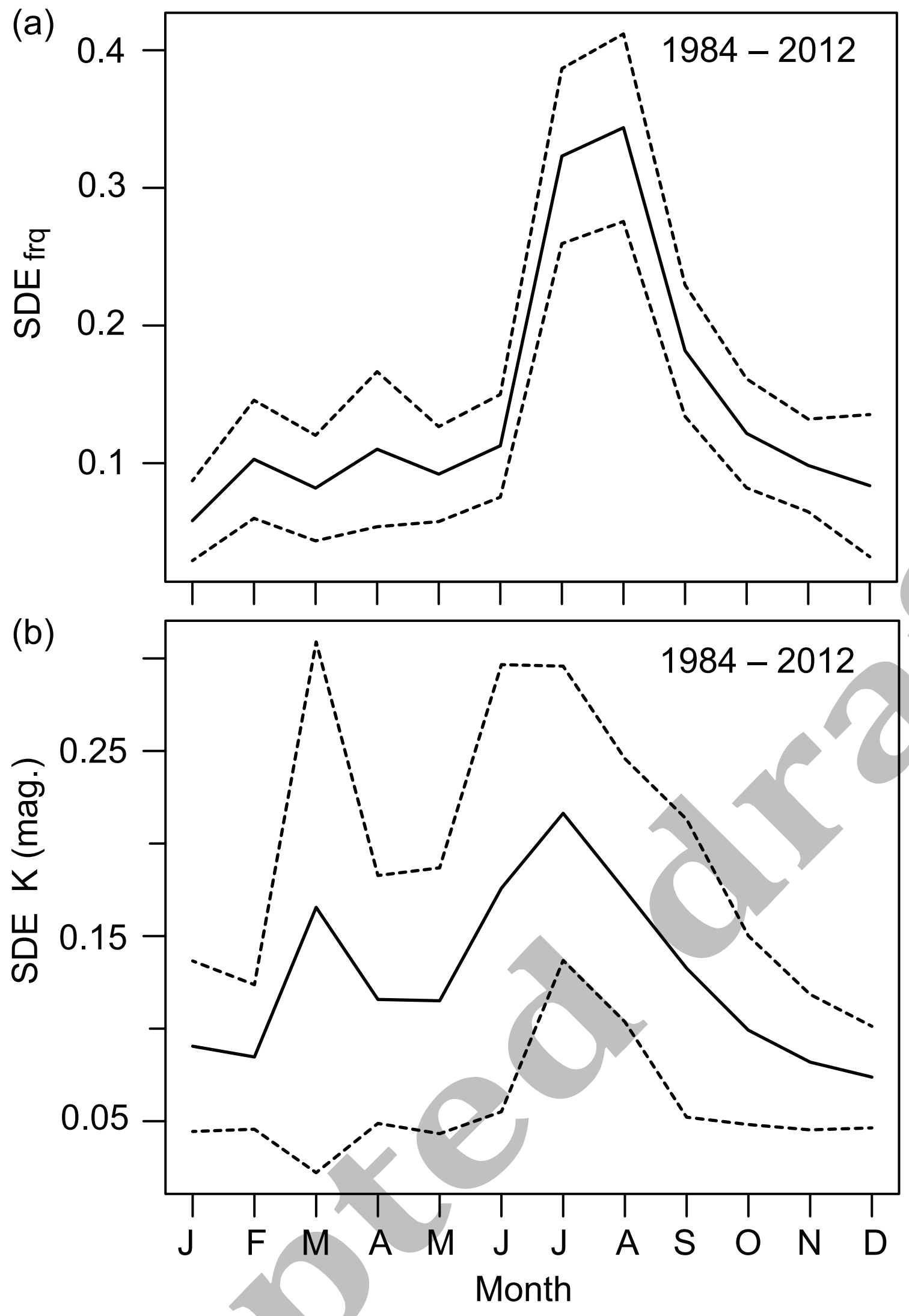

**Figure 5**. (a) Saharan Dust Event frequency ($SDE_{frq}$) per calendar month (number of nights above threshold extinction value divided by total number of observed nights), and (b) mean atmospheric extinction (K, units in magnitudes) of SDEs per calendar month over the 1984–2012 period. Dashed lines show ±1.96 standard error of the mean (SEM) confidence values.

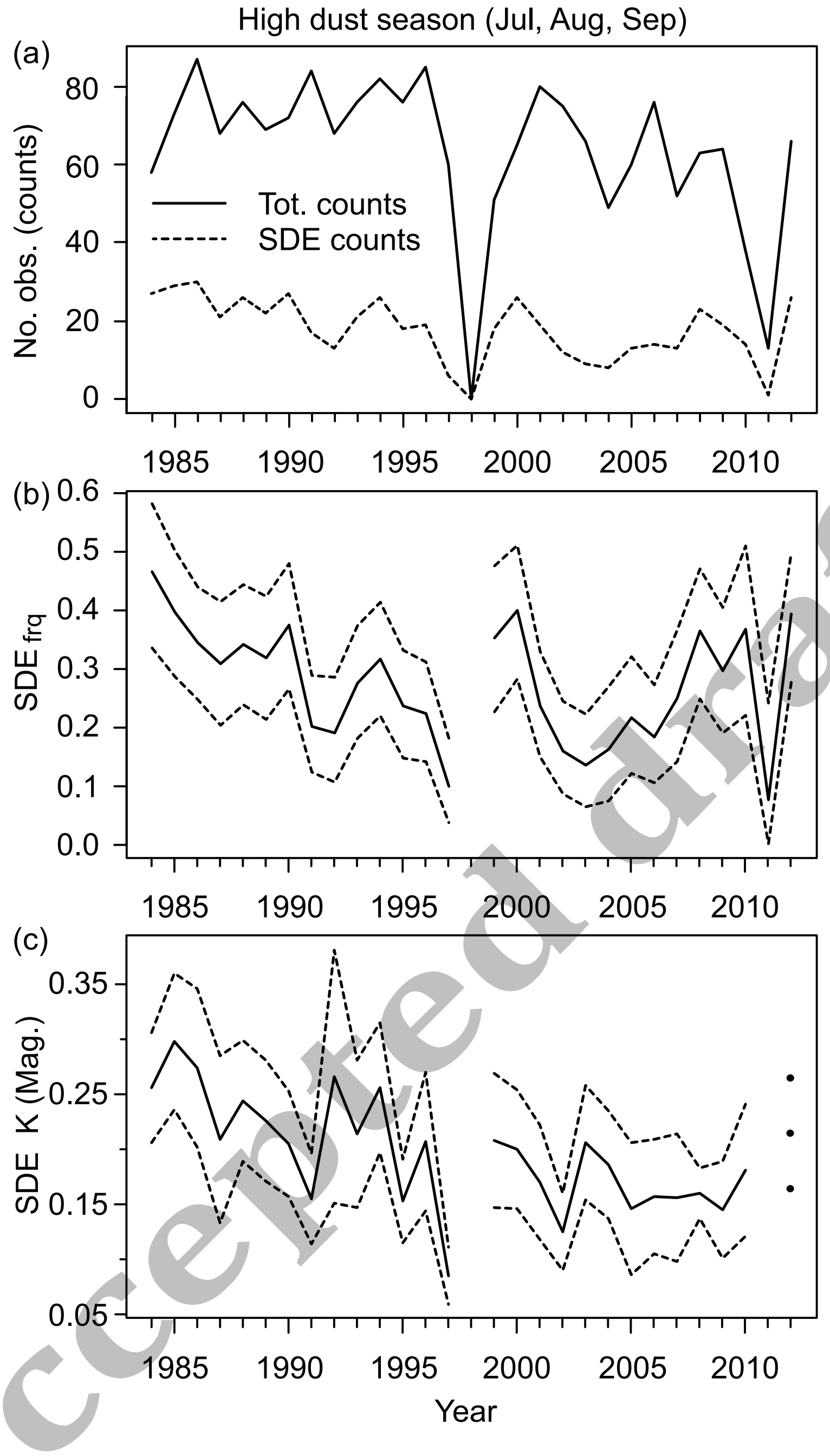

**Figure 6**. For the period of 01st July to 30th September (defined here as the high dust season), panels a–c respectively show over the years 1984–2012: (a) the total number of SDE observations (dashed line) and photometric observations (solid line) per-season; (b) the SDE frequency ($SDE_{frq}$), calculated from the number of nights with atmospheric extinction values observed above the identified threshold values, divided by the total number of observations per-season (solid line). The dashed lines show the two-tailed ±1.96σ level confidence intervals, calculated from binomial probability density estimates constrained by the number of SDEs/total observations. Panel (c) shows the mean atmospheric extinction (K, units in magnitudes) of all observed SDEs (solid line), with the ±1.96 standard error of the mean (SEM) confidence intervals shown on the dashed lines. For dust seasons with ≤1 observed SDE no data is displayed, such periods occur in 1998, and 2011: consequently, points are used to indicate the atmospheric extinction range of the 2012 season.

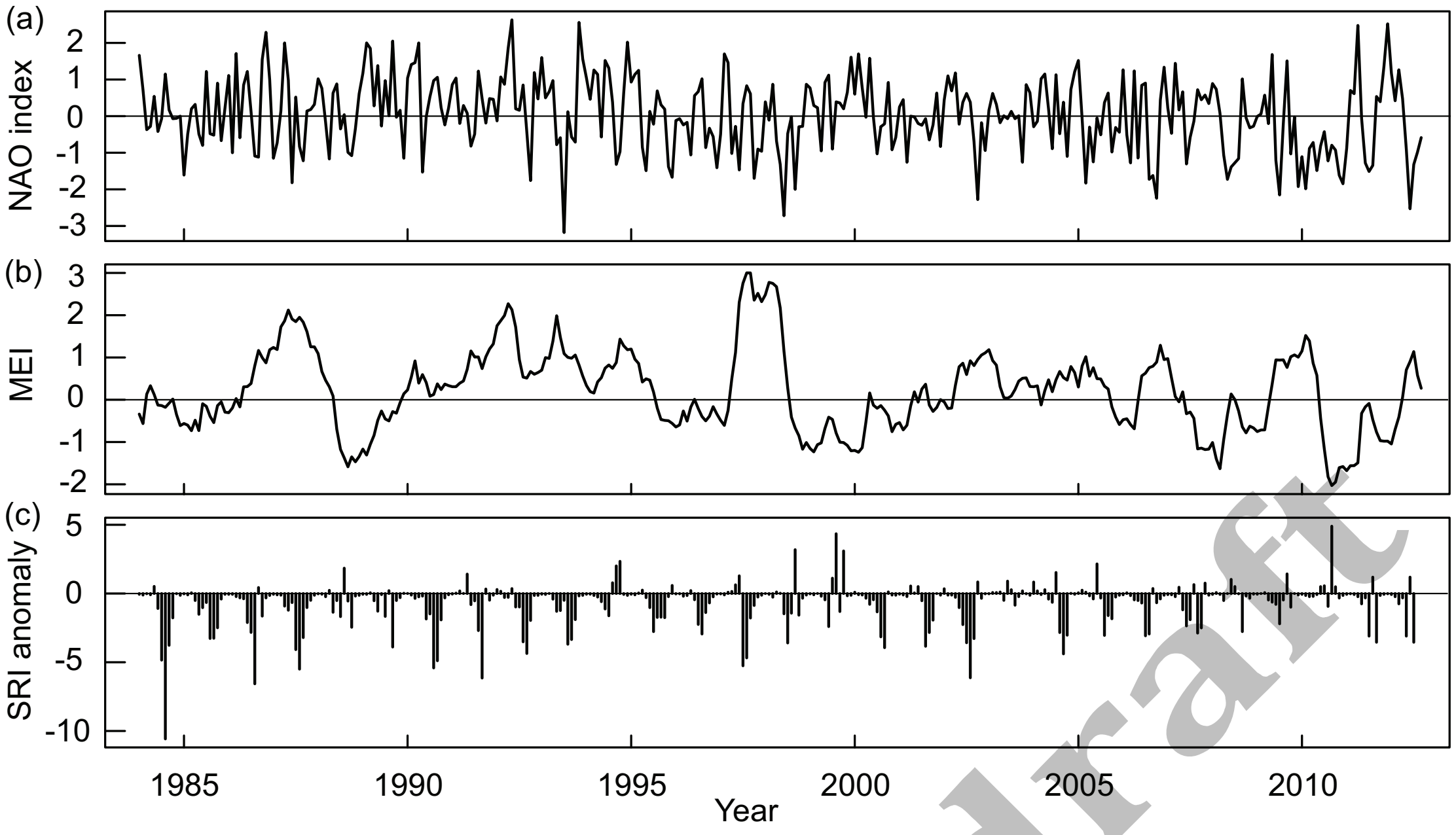

**Figure 7**. Monthly time series between 1984 to 2012 of (a) North Atlantic Oscillation (NAO) index, (b) Multivariate El Niño Southern oscillation index (MEI), and (c) Sahel Rainfall Index (SRI) anomaly (units in cm), where the anomalies are with respect to 1950–1979.

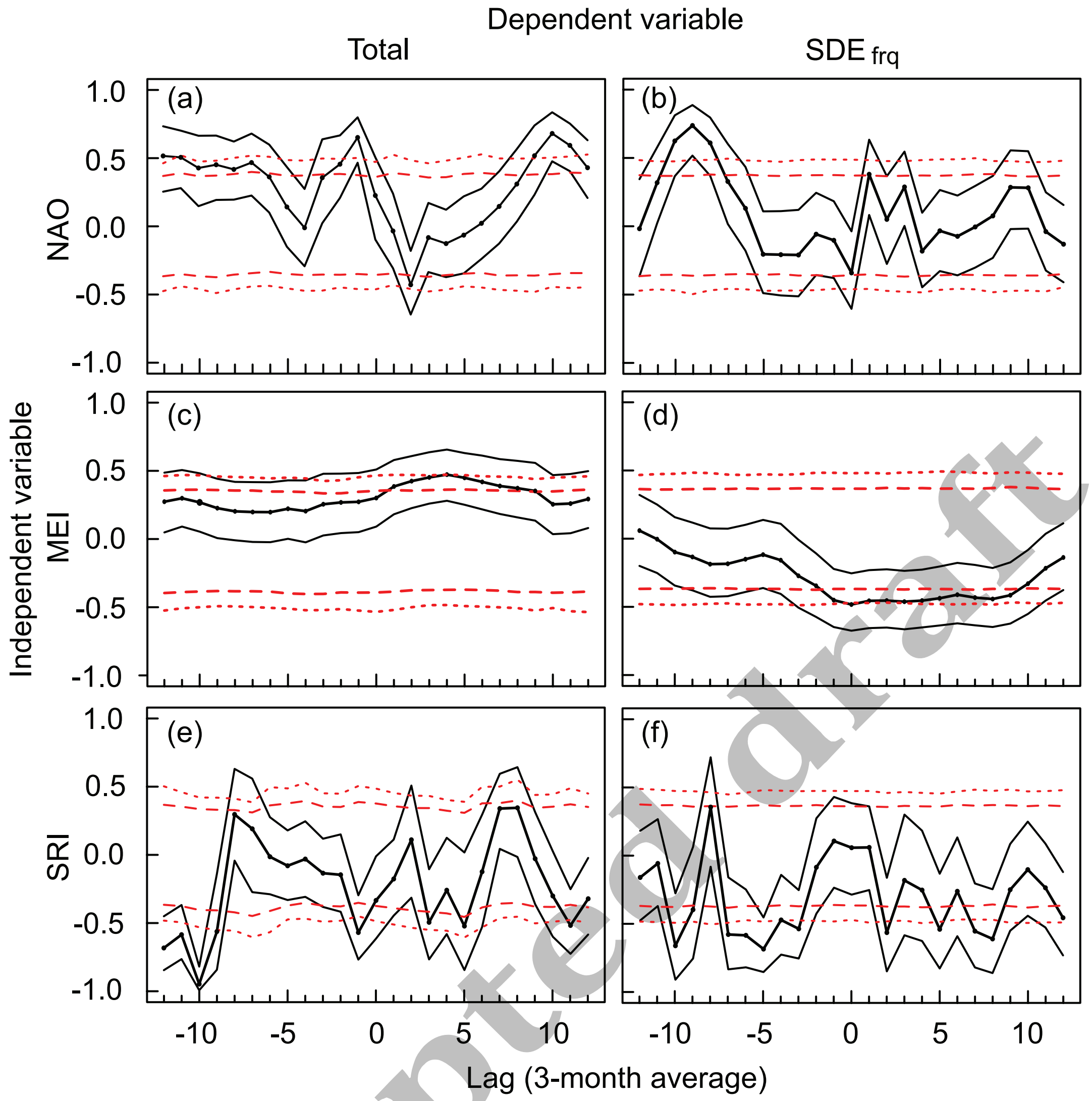

**Figure 8**. Lagged-correlations of the high dust season SDE properties (dependent variables) occurring over the three-month period of July–September (zero lag) for: Total (total number of observed nights), and SDE frequency ($SDE_{frq}$). These are compared to the independent variables: the NAO index, MEI, and SRI over a ±12 lag period, where, each *x*-axis value represents a three-month average with a delay corresponding to the lag value: i.e. a lag of -1 corresponds to values from the calendar months of June–August, whereas a lag of +1 corresponds to the months of August–October. Correlation coefficient distributions are calculated from Markov Chain Monte Carlo linear regressions which include a consideration of uncertainties in both the independent and dependent variables where available. The median and ±1σ intervals for these correlations are presented (solid black lines). The dashed and dotted red lines indicate the two-tailed 0.05 and 0.01 *p*-values respectively, calculated individually for each lag period from simulations of 10,000 correlations, where the independent variable is held constant while the dependent variable is randomized.

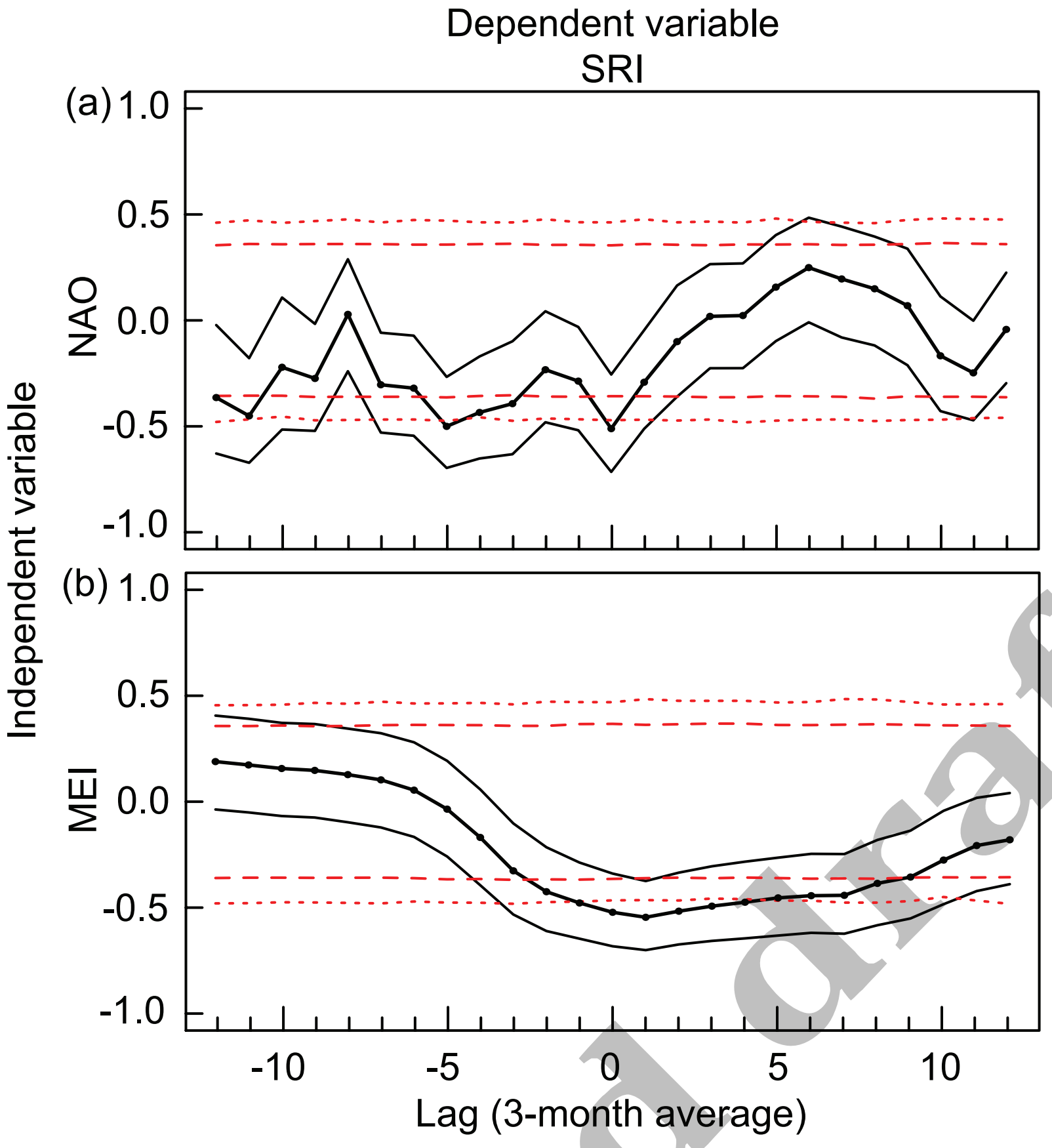

**Figure 9**. Identical to Figure 8, except the dependent variable now considered is the Sahel Rainfall Index (SRI), while the independent variables are the NAO and MEI.